\documentclass[
twocolumn,
aps,prd,
nofootinbib,
superscriptaddress,
showpacs,ligh
tightenlines,
]{revtex4}

\usepackage{amsmath}
\usepackage{amssymb}
\usepackage{slashed}

\usepackage{bm}
\usepackage{color,graphicx}

\usepackage{mciteplus}

\begin{document}

\title{Parton Transverse Momentum and Orbital Angular Momentum Distributions}

\author{Abha Rajan}
\email{ar5xc@virginia.edu}
\affiliation{University of Virginia - Physics Department,
382 McCormick Rd., Charlottesville, Virginia 22904 - USA} 

\author{Aurore Courtoy} 
\email{acourtoy@fis.cinvestav.mx}
\affiliation{Catedr\'atica CONACyT, Departamento de F\'isica, Centro de 
Investigaci\'on y de Estudios Avanzados, Apartado Postal 14-740, 07000 
M\'exico D.F., M\'exico}

\author{Michael Engelhardt}
\email{engel@nmsu.edu}
\affiliation{New Mexico State University - Department of Physics, Box 30001 MSC 3D, Las Cruces, NM 88003 - USA}

\author{Simonetta Liuti }
\email{sl4y@virginia.edu}
\affiliation{University of Virginia - Physics Department,
382 McCormick Rd., Charlottesville, Virginia 22904 - USA \\ and Laboratori Nazionali di Frascati, INFN, Frascati, Italy.}

\begin{abstract}
The quark orbital angular momentum component of proton spin, $L_q$, can be
defined in QCD as the integral of a Wigner phase space distribution
weighting the cross product of the quark's transverse position and momentum.
It can also be independently defined from the operator product expansion
for the off-forward Compton amplitude in terms of a twist-three generalized
parton distribution. We provide an explicit link between the two definitions,
connecting them through their dependence on partonic intrinsic transverse
momentum. Connecting the definitions provides the key for correlating direct
experimental determinations of $L_q$, and evaluations through Lattice QCD
calculations. The direct observation of quark orbital angular momentum does
not require transverse spin polarization, but can occur using longitudinally
polarized targets.
\end{abstract}

\maketitle
\baselineskip 3.0ex
Orbital Angular Momentum (OAM), $L_{q,g}$, is generated inside the proton
as a consequence of the quark and gluon transverse motion about the system's
center of momentum. 
It has been identified as a critical component in the resolution of
the proton spin puzzle \cite{Jaffe:1989jz}, which has constituted a
central focus of hadron physics since the seminal EMC experiments
demonstrated that quark spin alone
cannot account for the proton spin \cite{Ashman:1987hv,Ashman:1989ig}. 
%
Understanding OAM in the proton was the original motivation for
introducing Generalized Parton Distributions (GPDs) in
Refs.~\cite{Radyushkin:1997ki,Ji:1996ek}, in that they provided a novel
way of accessing angular momentum through a class of exclusive reactions
including Deeply Virtual Compton Scattering (DVCS), Deeply Virtual Meson
Production (DVMP), and related experiments. 
Through  Ji's sum rule \cite{Ji:1996ek}, one can, in fact, relate the
components of the Energy Momentum Tensor (EMT) known as the gravitomagnetic
form factors, $A_{q,g}$ and $B_{q,g}$, to the quark and gluon total angular
momenta, $J_{q,g}$. The pivotal observation made in \cite{Ji:1996ek} is
that $A_{q,g}$ and $B_{q,g}$ correspond to $n=2$ Mellin moments of
GPDs which, in turn, define the matrix elements for DVCS. These important
developments rendered total angular momentum a measurable quantity. 
Although the decomposition of $J_g$ into its spin and orbital components
has proven difficult to define gauge invariantly, the orbital angular
momentum of quarks is well defined through $J_q=L_q+S_q$. 
Even so, the direct observability of $L_q$ remains a challenging question:
the framework defined so far  does not tell us how to access the dynamics of quark orbital motion since $L_q$ is only obtained through the difference of the total angular momentum and spin components. 

$L_q$ has more recently been associated with precise operators and structure
functions, given within two alternative approaches. On one side, a dynamical
picture of quark orbital motion was given in terms of a Generalized
Transverse Momentum Distribution (GTMD), {\it i.e.}, an unintegrated over
transverse momentum GPD,
in Refs.~\cite{Burkardt:2008ua,Lorce:2011kd,Hatta:2011ku}.
%
%
The GTMD-based definition of quark OAM is
\begin{equation}
\label{Lu}
L^{\cal U}_q(x)  = \int d^2 k_T \int d^2 b_T \,  (b_T
\times k_T)_3 {\cal W}^{\, \cal U}(x, k_T, b_T )
\end{equation}
where ${\cal W}^{\, \cal U}$ is a Wigner distribution derived from
the quark-quark off-forward correlator in a longitudinally polarized nucleon
moving in the 3-direction\footnote{Throughout this paper we consider
zero skewness, i.e., the plus component of the momentum transfer vanishes,
$\Delta^{+} =0$. Moreover, we omit writing explicitly the $Q^2$ dependence
which is, however, present in all expressions.}
\begin{equation}
\Phi^{\Gamma }_{\Lambda^{\prime } \Lambda } (p^{\prime } ,p;
z^{\prime } ,z) = 
\langle p^{\prime } , \Lambda^{\prime } \mid \overline{\psi}(z^{\prime } )
\Gamma \, {\cal U} \, \psi(z) \mid p, \Lambda \rangle
\label{fundcorr}
\end{equation}
where $\Gamma $ denotes an arbitrary $\gamma $-matrix structure.
${\cal W}^{\, \cal U}$ is obtained by Fourier-transforming (\ref{fundcorr})
for $\Gamma = \gamma^{+} $ from $z-z^{\prime } $ to struck quark intrinsic
momentum $k$, projecting onto $(z-z^{\prime } )^{+} =0$,
as well as from (transverse) momentum transfer $\Delta_{T} $ to transverse
position $b_T $. If one foregoes the transformation to $b_T $,
one can relate $L^{\cal U}_q$ to the $k_T^2 $ moment of the GTMD $F_{14}$
\cite{Hatta:2011ku,Lorce:2011ni,Meissner:2009ww}
for $\Delta_{T} \rightarrow 0$,
\begin{eqnarray}
\label{F14_1}
L^{\cal U}_q (x) = - F_{14}^{(1)} & \equiv &\!\!\!  - \int d^2 k_T \,
\frac{{k}_T^2}{M^2} F_{14}(x, {k}_T^2, {k}_T \cdot \Delta_{T}, \Delta_{T}^{2} )
\,.\nonumber\\
\end{eqnarray}
$F_{14}$ is a GTMD describing an unpolarized quark inside a longitudinally
polarized proton \cite{Meissner:2009ww}. Finally, ${\cal U}$ in
Eq.~(\ref{fundcorr}) denotes the gauge link, {\it i.e.}, the Wilson
path-ordered exponential connecting the coordinates $z$ and $z^{\prime }$.
We will restrict the discussion in the present Letter to the case of a
straight gauge link, corresponding to what is known as Ji's decomposition of 
angular momentum \cite{Burkardt:2012sd}, and defer the analogous treatment
of other relevant gauge link structures to an expanded exposition.

In another approach \cite{Penttinen:2000dg,Kiptily:2002nx,Hatta:2012cs},
it was observed that OAM is associated with a twist-three GPD, $G_2 $.
Similar to the treatment of the forward case \cite{Mulders:1995dh,Boer:1997bw,Bacchetta:2006tn}, one can write the
Mellin moments of $G_2$, which appears in the parametrization of the
off-forward amplitude, in terms of both twist-two operators and (genuine)
twist-three operators. For the second moment, the genuine twist-three
contribution vanishes and one obtains, for $\Delta_{T} \rightarrow 0$,
\begin{eqnarray}
\label{polyakov}
\int_0^1 dx  \, x G_2 & = & -\frac{1}{2} \int_0^1 dx\, x (H+E) +
\frac{1}{2}  \int_0^1 dx \, \tilde{H} \nonumber \\
& = & -J_q + S_q = -L_q^{\rm Ji}
\end{eqnarray}
where only a straight gauge link structure applies in such a relation
involving only GPDs.
This result can be viewed as an extension of the Efremov-Leader-Teryaev
(ELT) sum rule \cite{Efremov:1996hd}, written for the polarized structure
functions, to off-forward kinematics.

Notwithstanding these developments, two main problems remain to be solved: 1) relating the two distinct structures, one ($F_{14}$) appearing at twist two, and one ($G_2$) at twist three, both describing OAM within the same gauge invariant framework;   2) singling out an experimental measurement to access directly OAM, possibly through the newly defined structures.
In this Letter, we provide a direct link between the $k_T^2 $ moment of the GTMD
and the twist-three GPD describing OAM, elucidating the underlying dependence
on partonic intrinsic transverse momentum and off-shellness. 
The GTMD-based definition is
calculable in Lattice QCD using the techniques of Ref.~\cite{Musch:2011er}.
On the other side, the twist-three GPD-based definition can be measured
directly in DVCS-type experiments, through 
the azimuthal angle modulations which are sensitive to twist-three GPDs
in DVCS off a longitudinally polarized target \cite{Courtoy:2013oaa};
this is at variance with the notion that transverse polarization,
or proton spin-flip processes are necessary to obtain information on quark OAM.

Our central result is the following integral relations
(\ref{relation2},\ref{relation1}) connecting $F_{14}$, $G_2$,
$\widetilde{E}_{2T} $, $H$, $E$ and $\widetilde{H} $ in the limit
$\Delta_{T} \rightarrow 0$, where
$\widetilde{E}_{2T} $ is a twist-three GPD in the classification
of \cite{Meissner:2009ww} related to the GPD $G_2$ in the classification
of \cite{Kiptily:2002nx} by
\begin{equation}
\int dx\, x \widetilde{E}_{2T}  = -\int dx\, x ( H +E +  G_2) \ ,
\label{g2connect}
\end{equation}

\begin{widetext}
\begin{eqnarray}
{\rm (LIR)} \ \ \ \ \ \ \ \ \ \ \ \ \ \ \ \ \ \ \ \ F_{14}^{(1)} &=&
-\int\limits_{x}^{1} dy\, \left(\widetilde{E}_{2T} + H + E\right)
\ \ \ \ \Rightarrow \ \ \ \
-L_q^{\rm Ji} = \int\limits_{0}^{1} dx
F_{14}^{(1)} = \int\limits_{0}^{1}  dx\, x \, G_2
\label{relation2} \\
{\rm (EoM)} \;\;\;\; x (\widetilde{E}_{2T} + H + E)
&=& x \left[ (H+E)
- \int\limits_{x}^1 \frac{dy}{y} (H+E) - \frac{1}{x} \widetilde{H}
+ \int \limits_{x}^1 \frac{dy}{y^2}  \, \widetilde{H} \right] + G^{(3)}
= x(\widetilde{E}_{2T} + H + E)^{WW} + G^{(3)} \nonumber \\
\label{relation1}
\end{eqnarray}
\end{widetext}
Eq.~(\ref{relation2}) is a Lorentz Invariance Relation (LIR), obtained from
the analysis of the most general Lorentz decomposition of the quark-quark
correlation function. It states a remarkable equivalence between the GTMD
and twist-three GPD-based definitions.  

\noindent Eq.~(\ref{relation1}), which integrates over $x$ to give (\ref{polyakov}),
is an Equations of Motion (EoM) relation derived by
applying the QCD EoM for the quark fields to the correlation function.
Together with Eq.~(\ref{relation2}), it allows one to connect OAM defined
through a Wigner distribution, Eq.~(\ref{F14_1}), to the sum rule
definition, Eq.~(\ref{polyakov}). In Eq.~(\ref{relation1}), the superscript
``$WW$'' denotes the Wandzura-Wilczek part, in analogy to the derivation for the
polarized structure functions $g_1$ and $g_2$ \cite{Wandzura:1977qf}.
On the other hand,
\begin{equation}
G^{(3)} = -\widetilde{\cal M}
+x\int_{x}^{1} \frac{dy}{y^2 } \widetilde{\cal M}
\end{equation}
with $\widetilde{\cal M} $ given in Eq.~(\ref{gen3}) below,
is a genuine twist-three term -- a quark-gluon-quark correlation
-- whose contribution to angular momentum can be explicitly seen to
vanish in the case treated here as a consequence of the underlying
structure of Eq.~(\ref{covder}).

We now sketch the derivation of Eqs.(\ref{relation2},\ref{relation1}),
highlighting the role of quark $k_T$ and, thus, the off-shellness of
partons in generating proton spin. The completely unintegrated
off-forward quark-quark correlation function $W^\Gamma_{\Lambda,\Lambda'}$,
{\it i.e.}, (half) the four-dimensional Fourier transform of (\ref{fundcorr})
from $z-z^{\prime } $ to $k$
\cite{Mulders:1995dh,Boer:1997bw,Bacchetta:2006tn,Goeke:2005hb,Meissner:2009ww},
can be parametrized \cite{Meissner:2009ww} in terms of invariant
functions $A_i $. On the other hand, its $k^{-} $ integral
$\widetilde{W}^{\Gamma }_{\Lambda^{\prime } ,\Lambda } =\int dk^{-}
W^{\Gamma }_{\Lambda^{\prime } ,\Lambda } $ is parametrized by the GTMDs.
This implies the following twist-two relations already given in
\cite{Meissner:2009ww}, adapted to the straight gauge link, zero skewness
case considered here,
\begin{eqnarray}
\frac{k_T \cdot \Delta_{T} }{\Delta_{T}^{2} } F_{12} +F_{13} &=&
2P^{+} \int dk^{-} \left(
\frac{k_T \cdot \Delta_{T} }{\Delta_{T}^{2} } A_5 + A_6 \right. \nonumber \\
& & \left. -\frac{xP^2 - k\cdot P }{M^2 } (A_8 +xA_9 ) \right) \label{f123} \\
F_{14} &=& 2P^{+} \int dk^{-} \left( A_8 +xA_9 \right)
\end{eqnarray}
which we supplement by the twist-three relation
\begin{eqnarray}
\frac{k_T \cdot \Delta_{T} }{\Delta_{T}^{2} } F_{27} +F_{28} =
2P^{+} \int dk^{-} \left( \frac{k_T \cdot \Delta_{T} }{\Delta_{T}^{2} } A_5
+ A_6 \right. & & \nonumber \\
+\frac{1}{M^2 } \left. \left(
\frac{(k_T \cdot \Delta_{T} )^2 }{\Delta_{T}^{2} } -k_T^2 \right)
A_9 \right) \ . \ \ \ & & \label{f278}
\end{eqnarray}
Combining integrals over transverse $k_T $ of these relations,
one arrives at the LIR
\begin{equation}
\label{LIR2_alt}
\frac{d}{dx}  \int d^2 k_T  \, \frac{ k_T^2}{M^2} \, F_{14} =
\widetilde{E}_{2T} + H + E
\end{equation}
in the limit $\Delta_{T} \rightarrow 0$,
having identified the GPD combinations $H+E$ and $\widetilde{E}_{2T} $
resulting after $k_T $ integration of the GTMD combinations appearing
in (\ref{f123}) and (\ref{f278}) \cite{Meissner:2009ww}.
Finally, integrating over $x$, one arrives at Eq.~(\ref{relation2}).

The EoM relation in Eq.~(\ref{relation1}) was obtained by considering
(\ref{fundcorr}) for $\Gamma=i \sigma^{i+} \gamma_5$, $(i=1,2)$,
and inserting the equation of motion for the quark operator (the
symmetrized form serving to cancel the mass terms),
\begin{eqnarray}
& 0 & = \displaystyle\int \frac{d z^- d^2 z_T}{(2 \pi)^3}
e^{ixP^+ z^- - ik_T\cdot z_T} \times \\
& &  \!\!\!\! \langle p', \Lambda' \mid \overline{\psi}(-z/2)
(\Gamma \, {\cal U} \, i\overrightarrow{\slashed{D} }
+ i\overleftarrow{\slashed{D} } \, \Gamma \, {\cal U} )
\psi(z/2) \!  \mid p, \Lambda \rangle_{z^+=0} .
\nonumber
\end{eqnarray}
This yields the relation between $k^{-} $ integrated correlators
\begin{equation}
\label{intermed}
-xP^+ i \epsilon_T^{ij} \widetilde{W}_{\Lambda \Lambda'}^{\gamma^j} =
\frac{\Delta_{T}^{i} }{2}
\widetilde{W}_{\Lambda \Lambda'}^{\gamma^+ \gamma_5} -
k_T^j i \epsilon_T^{ij} \widetilde{W}_{\Lambda \Lambda'}^{\gamma^+}  
+ {\cal M}^i_{\Lambda \Lambda'}\ , 
\end{equation}
with the genuine twist-three quark-gluon-quark correlator
(still denoting $\Gamma=i \sigma^{i+} \gamma_5$),
\begin{widetext}
\begin{eqnarray} 
{\cal M}^i_{\Lambda \Lambda'} &=& \frac{1}{4}
\int \frac{d z^- d^2 z_T}{(2 \pi)^3} e^{ixP^+ z^- - i k_T\cdot z_T} 
\langle p^{\prime } ,\Lambda^{\prime } \mid \overline{\psi}(-z/2) \left[
\left. (\overrightarrow{\slashed{\partial } } -ig\slashed{A} )
{\cal U} \Gamma \right|_{-z/2} + \left. \Gamma {\cal U}
(\overleftarrow{\slashed{\partial } } +ig\slashed{A} ) \right|_{z/2}
\right] \psi (z/2)  \mid p, \Lambda \rangle_{z^+=0} \nonumber \\
& & \label{covder}
\end{eqnarray}
\end{widetext}
By taking the proton non-flip spin components,
$(\Lambda,\Lambda')= (+,+) - (-,-)$ that identify OAM \cite{Courtoy:2013oaa},
using the GTMD parametrizations \cite{Meissner:2009ww} of the
$\widetilde{W}^{\Gamma }_{\Lambda^{\prime } ,\Lambda } $, and
integrating over $k_T$, one has, in the $\Delta_{T} \rightarrow 0 $ limit,
\begin{equation}
\label{intermed2}
-x \widetilde{E}_{2T} = \widetilde{H} -\int d^2 k_T \,
\frac{k_T^2 }{M^2} F_{14} + \widetilde{\cal M}
\end{equation}
having again identified the GPD $\widetilde{E}_{2T} $ as in the LIR
derivation above, as well as the GPD
$\widetilde{H} = \int d^2 k_T \, G_{14} $.
The genuine twist-three term $\widetilde{\cal M} $ is given by
\begin{equation}
\widetilde{\cal M}
 = 2M \frac{\Delta_{T}^{i} }{\Delta_{T}^{2} }
\int d^2 k_T \left[ {\cal M}^i_{++} - {\cal M}^i_{- - }  \right]\,. 
\label{gen3}
\end{equation}
The final expression defining the EoM
relation in Eq.~(\ref{relation1}) is obtained by taking the derivative in
$x$, and replacing the $k_T^2 $ moment of $F_{14}$ with the expression from
the LIR in Eq.~(\ref{relation2}).

It should be noted that the relations discussed here are perturbatively
divergent and require consistent regularization/renormalization at each
step. An interesting aspect, e.g., of the LIR (\ref{relation2}) is that
it connects a GTMD which does not have a GPD limit, $F_{14} $, to GPDs.
To treat both sides on an equal footing implies utilizing a
transverse momentum-dependent regularization and renormalization scheme,
and thus interpreting the GPDs in terms of the underlying GTMDs of which they
are the GPD limit. On the other hand, it seems tempting to speculate that
relations of the type (\ref{relation2}) may ultimately be useful to
connect the renormalization of quantities which are intrinsically
defined as transverse momentum-dependent, such as $F_{14} $, to the
more standard schemes employed for GPDs.

\begin{figure}
\includegraphics[width=9cm]{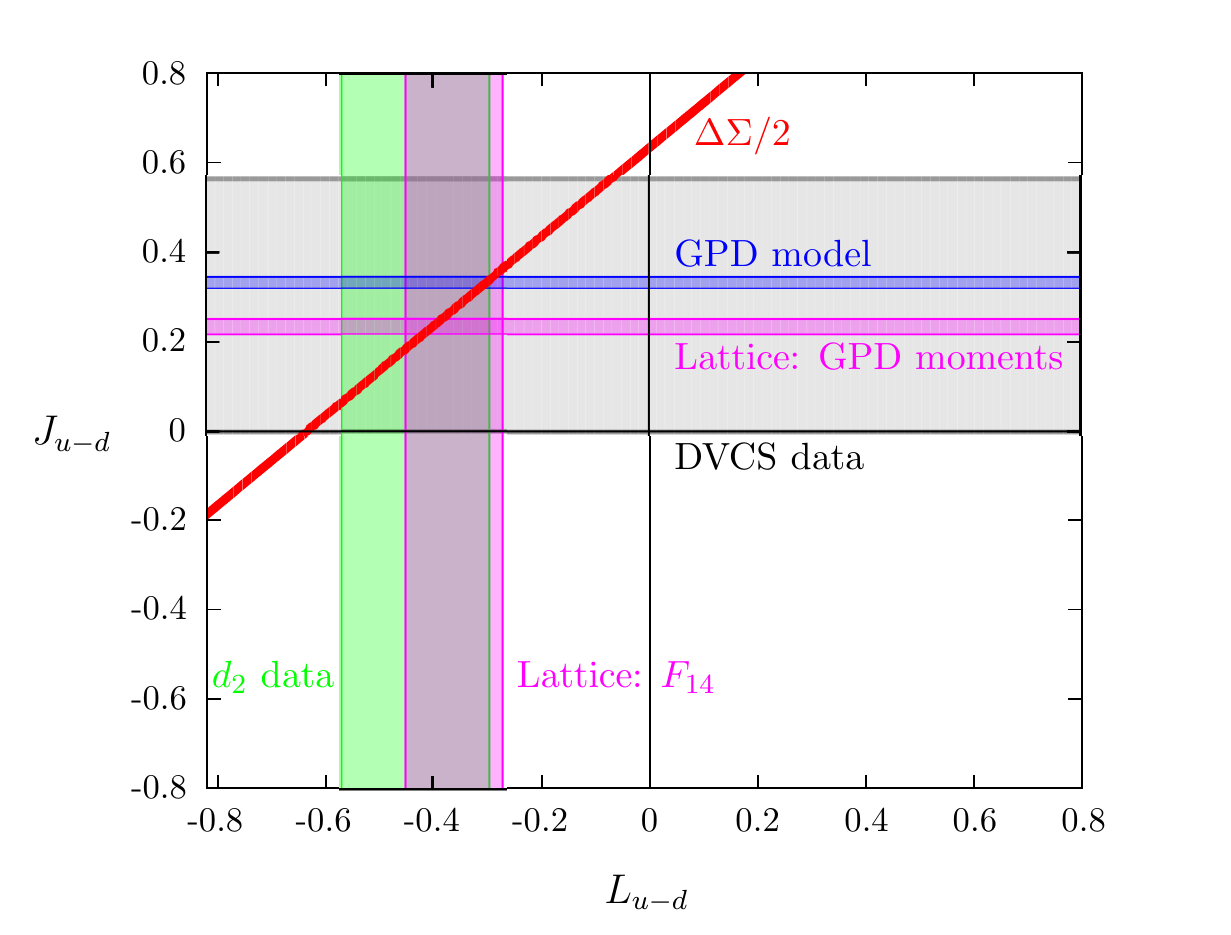}
\caption{$J_{u-d}$ plotted vs. $L_{u-d}$. The (red) slanted band represents
$J_{u-d}= L_{u-d}+(1/2) \Delta \Sigma_{u-d}$ using $\Delta \Sigma$
from Ref.~\cite{Airapetian:2006vy}. The horizontal bands represent
$J_{u-d}$ from experiment (gray) \cite{Mazouz:2007aa}, from the GPD model
extraction (blue) \cite{Goldstein:2010gu,GonzalezHernandez:2012jv}, and
from lattice QCD (magenta) \cite{Bratt:2010jn}. The vertical bands are
the preliminary lattice QCD evaluation of $L_{u-d}$ using the
definition in Eq.~(\ref{Lu}) (magenta) \cite{poetic:engelhardt},
and the GPD model normalized according to Eq.~(\ref{d2G2}) (green).}
\label{fig1}
\end{figure}
As an application of the relations between the different ways to access
angular momentum, we compile and correlate in Fig.~\ref{fig1}
determinations of $J_q $, $L_q $ and $S_q $ from several sources,
including experiment, lattice QCD, and models.
The value of $J_{u-d}=J_u-J_d$ is plotted versus $L_{u-d}=L_u-L_d$.
The horizontal bands represent measurements/calculations of $J_{u-d}$
using DVCS data \cite{Mazouz:2007aa}/GPD evaluations; the slanted band
is given by the relation $J_q = L_q + \Delta \Sigma_q /2$, where the
experimental value for $\Delta \Sigma_{u-d}$ was taken from
Ref.~\cite{Airapetian:2006vy}.
The vertical bands correspond to preliminary data for $L_q$ obtained
in a lattice QCD calculation at an artificially high pion mass of
$m_{\pi } =518\, \mbox{MeV} $ using an approach related to the GTMD
$F_{14}$ from Eq.~(\ref{F14_1}) \cite{poetic:engelhardt}, and to a
calculation of $\widetilde{E}_{2T} $ in the reggeized
diquark model 
\cite{Goldstein:2010gu,GonzalezHernandez:2012jv}.
The lattice result is expected to be enhanced by roughly 30\%
as one goes to the physical pion mass. The reggeized diquark model
produces a parametrization of the GPDs 
$H$ and $E$, which is fitted to both the nucleon unpolarized PDFs for
the $u$ and $d$ quarks, and to the flavor-separated nucleon electromagnetic
form factors \cite{Cates:2011pz}. An independent experimental constraint
on the normalization of the genuine twist-three part of $\widetilde{E}_{2T} $
is obtained by using its third Mellin moment, which can be related to
\begin{eqnarray}
\label{d2G2}
d_2 = 3\int_0^1 dx x^2 g_2^{tw3} (x) \ ,
\end{eqnarray}
where $g_2$, the transverse spin-dependent structure function, 
is obtained in double-spin asymmetry measurements of longitudinally
polarized electrons scattering from longitudinally and transversely
polarized nucleons. We used, in particular, the SLAC data for the $u$
and $d$ quark values of $d_2$ at a common $Q^2$ value of 5 GeV$^2$
\cite{Anthony:2002hy}. With the normalization of the twist-three part of
$\widetilde{E}_{2T} $ obtained from Eq.~(\ref{d2G2}) we then evaluated
$L_q$. The result is the vertical green band. This is consistent, although
with a large error, with the values extracted from the lattice. No
experimental determinations of $L_q$ to corroborate our analysis can be
placed on the graph at this point, although future extractions will be
possible from analyses of the $\sin 2 \phi$ modulation of DVCS data
\cite{Courtoy:2013oaa}.

\begin{figure}
\includegraphics[width=8cm]{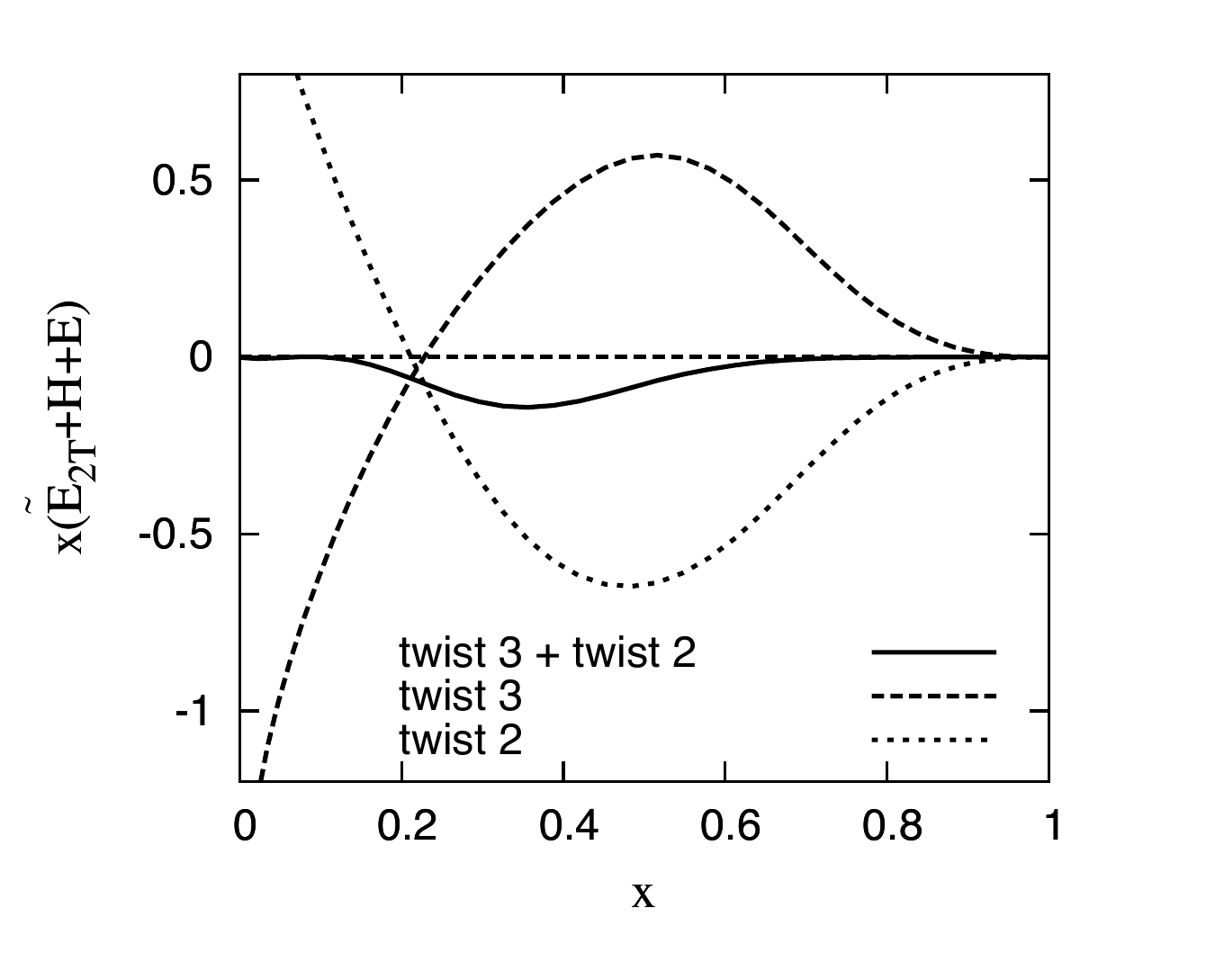}
\caption{Contributions in Eq.~(\ref{relation1}) calculated in the
reggeized diquark model \cite{GonzalezHernandez:2012jv}. The dashed
line is the genuine twist-three contribution, the dotted line is the
twist-two term, $x(\widetilde{E}_{2T} + H + E)^{WW}$, and the full line
is their sum. All quantities are evaluated for $\Delta_{T}=0$ at the initial scale of the model.}
\label{fig3}
\end{figure}
In Fig.~\ref{fig3} we exhibit in more detail the contributions in
Eq.~(\ref{relation1}) as a function of $x$, i.e., the behavior of
$x(\widetilde{E}_{2T} + H + E)^{WW} $, the genuine twist-three term, and
their sum at the initial scale of the model.
As for $g_2$, the genuine twist-three part is predicted to be large.
Due to the Regge behavior of the functions, we expect measurements at
low $x$, {\it i.e.}, in a regime which would be best accessible at an Electron Ion Collider to be important.

Finally, future developments will include the extension of our study to
the Jaffe-Manohar \cite{Jaffe:1989jz} decomposition of angular momentum,
which, as shown in Ref.~\cite{Burkardt:2012sd}, involves a final state
interaction (encoded in a staple-shaped gauge link), and is related to
the Ji decomposition by
\begin{eqnarray}
\label{Burkardt:eq1}
L^{\rm{JM} }_q  =  L^{\rm{Ji} }_q  + \langle \tau_{3}  \rangle
\end{eqnarray}
where $\langle \tau_{3} \rangle$
is an off-forward extension of a Qiu-Sterman term \cite{Qiu:1991pp}.
$ \langle \tau_{3} \rangle$
has been interpreted physically as a change in OAM due to a torque - a
final state interaction - exerted on the outgoing quark by the
color-magnetic field produced by the spectators \cite{Burkardt:2012sd}. 


In conclusion, understanding quark OAM entails cross-correlating
phenomenology, theory and lattice QCD efforts to bring them to bear
simultaneously on the subject. We provided relations that are key
for realizing such a coordinated approach, utilizing directly
non-local, $k_T$-unintegrated quark-quark correlation functions.
This approach opens up an avenue to explore the role of partonic
transverse momentum and off-shellness for OAM, while providing a
formalism which connects to lattice QCD calculations on one side
and to experiment on the other. A first, exploratory direct calculation
of quark OAM in lattice QCD using an approach related to the GTMD $F_{14}$
was incorporated into the analysis, and confronted with independent
determinations, e.g., via Ji's sum rule, and through $d_2$ measurements.
Our relations bring to the fore the intricacies of connecting a
twist-two GTMD moment and a twist-three GPD, before the backdrop of
a field theoretic rendition of OAM.

{\it Acknowledgments}: Discussions with M.~Burkardt, M.~Diehl, G. Goldstein, A.~Klein , S. Pate and the Theory Group at Jefferson Lab are
gratefully acknowledged. This work was supported by the U.S.~DOE and
the Office of Nuclear Physics through grants DE-FG02-96ER40965 (M.E.),
DE-AC05-06OR23177, and DE-FG02-01ER41200 (A.R. and S.L.).

\bibliography{OAM_bib}

\end{document}